\documentclass[
 reprint,
 superscriptaddress,
 amsmath,amssymb,
 aps,
 prl
]{revtex4-2}

\usepackage{graphicx}
\usepackage{dcolumn}
\usepackage{bm}
\usepackage[hidelinks,colorlinks=true,linkcolor=blue,citecolor=blue, anchorcolor = blue, urlcolor = blue]{hyperref}
\usepackage[mathlines]{lineno}

\usepackage{braket}
\usepackage{color}
\usepackage{float}

\newcommand{\afftsing}{\affiliation{State Key Laboratory of Low Dimensional Quantum Physics, Department of Physics, Tsinghua University, Beijing 100084, China}}
\newcommand{\affsiqse}{\affiliation{Shenzhen Institute for Quantum Science and Engineering, Southern University of Science and Technology, Shenzhen 518055, China}}
\newcommand{\affiqa}{\affiliation{International Quantum Academy, Shenzhen 518048, China}}
\newcommand{\affbqis}{\affiliation{Beijing Academy of Quantum Information Sciences, Beijing 100193, China}}
\newcommand{\affc}
{\affiliation{Hefei National Laboratory, Hefei 230088, P. R. China}}
\newcommand{\afffscqi}{\affiliation{Frontier Science Center for Quantum Information, Beijing 100084, People’s Republic of China}}

\newcommand{\pauli}[1]{\sigma_{\mathrm{#1}}}
\newcommand{\yb}{^{171} \mathrm{Yb}^+}
\newcommand{\figref}[2]{Fig.~\ref{#1}\textcolor{blue}{#2}}
\newcommand{\exfigref}[2]{Extended Data Fig.~\ref{#1}\textcolor{blue}{#2}}
\newcommand{\equref}[1]{Equation~(\ref{#1})}

\newcommand{\ce}{\mathrm{e}}
\newcommand{\ci}{\mathrm{i}}

\begin{document}
\title{Realization of programmable Ising models in a trapped-ion quantum simulator}

\afftsing
\affsiqse
\affiqa
\affbqis

\author{Yao Lu}
 \email{luy7@sustech.edu.cn}
 \affsiqse
 \affiqa
 \afftsing
\author{Wentao Chen}
 \afftsing
\author{Shuaining Zhang}
 \afftsing
\author{Kuan Zhang}
 \afftsing
\author{Jialiang Zhang}
 \afftsing
\author{Jing-Ning Zhang}
 \affbqis
\author{Kihwan Kim}     
 \email{kimkihwan@mail.tsinghua.edu.cn}
 \afftsing
 \affbqis
 \affc
 \afffscqi

\date{\today}

\begin{abstract}
A promising paradigm of quantum computing for achieving practical quantum advantages is quantum annealing or quantum approximate optimization algorithm, where the classical problems are encoded in Ising interactions. However, it is challenging to build a quantum system that can efficiently map any structured problems. Here, we present a programmable trapped-ion quantum simulator of an Ising model with all-to-all connectivity with up to four spins. We implement the spin-spin interactions by using the coupling of trapped ions to multiple collective motional modes and realize the programmability through phase modulation of the Raman laser beams that are individually addressed on ions. As an example, we realize several Ising lattices with different interaction connectivities, where the interactions can be ferromagnetic or anti-ferromagnetic. We confirm the programmed interaction geometry by observing the ground states of the corresponding models through quantum state tomography. Our experimental demonstrations serve as an important basis for realizing practical quantum advantages with trapped ions.
\end{abstract}

\maketitle

\section{Introduction} \label{sec:intro}
Significant theoretical and experimental progress has been made toward realizing practical quantum applications on Noisy Intermediate-Scale Quantum (NISQ) processors, which could lead to surpassing classical computational capabilities \cite{preskill2018quantum,moll2018quantum,bharti2022noisy}. Quantum annealing (QA) \cite{kadowaki1998quantum,farhi2001quantum,albash2018adiabatic,hauke2020perspectives,yarkoni2022quantum} and quantum approximate optimization algorithm (QAOA) \cite{farhi2014quantum,hadfield2019quantum,zhou2020quantum,sack2021quantum} are representative algorithms that can bring out quantum advantages in the application of practical problems such as combinatorial optimization, quantum many-body physics, and quantum chemistry. In these algorithms, a problem of interest is mapped to a quantum Hamiltonian, where the solution of the problem can be found by either adiabatic evolution to the ground state of the Hamiltonian \cite{farhi2001quantum} or time evolution under alternating phase and mixing Hamiltonians with the assistance of classical parameter search \cite{farhi2014quantum}. It has not yet clearly shown the possibility of surpassing classical computational power with the quantum processors in the NISQ regime \cite{ronnow2014defining,preskill2018quantum,bharti2022noisy}. Currently, many studies are being conducted heuristically to find quantum speed-ups with existing quantum processors \cite{johnson2011quantum,boixo2014evidence,king2022coherent,barends2016digitized,pagano2020quantum,mohseni2022ising,ebadi2022quantum,zhu2022multi}. 

However, due to system limitations, many quantum processors are still unable to fully map classically hard problems. For example, when a combinatorial optimization problem is encoded to an Ising model, a fully connected network with programmability is required for general mapping, which has been a major challenge in experimental development. It has been proposed to realize all-to-all connectivity from local interactions at the expense
of quadratic enlarge of the qubit numbers \cite{choi2008minor,lechner2015quantum}. For Rydberg atoms, it was proposed to use chains of additional atoms, called quantum wires to implement programmable all-to-all connectivity \cite{qiu2020programmable}, and an alternative quantum wire scheme, called the Rydberg quantum wire, with much less demanding experimental requirements, has been proposed and demonstrated \cite{kim2022rydberg}. On the one hand, ion-trap systems contain naturally all-to-all connectivity \cite{murali2019full}, which has been shown through the realization of global entangling gates and parallel gates \cite{lu2019global,figgatt2019parallel}. It also has been theoretically proposed to realize the two-dimensional connectivity of Ising models with a linear chain of ions \cite{korenblit2012quantum,shapira2020theory,manovitz2020quantum} and a chiral dynamics with qubits in a magnetic field gradient has been experimentally demonstrated \cite{shapira2023quantum}.  

In this paper, we implement arbitrarily programmable interactions of Ising models and prepare the ground states of the corresponding Hamiltonians through adiabatic evolution using up to four trapped-ion qubits. We realize the adiabatic evolution digitally \cite{mezzacapo2014digital,barends2016digitized,parra-rodriguez2020digital-analog}, where the Hamiltonians for spin-spin interactions and transverse fields are alternatively applied. The programmable spin-spin interactions are implemented by coupling trapped ions to multiple collective motional modes through mainly phase modulation of the individually addressing Raman laser beams \cite{haffner2008quantum,blatt2012quantum,schneider2012experimental,monroe2021programmable,lu2019global,figgatt2019parallel,wang2022fast}. The digital simulation suppresses the residual qubit-motion entanglements that introduce decoherence during the adiabatic evolution \cite{wang2012intrinsic,wang2013phonon,dylewsky2016nonperturbative,wall2017boson}. We confirm the prepared ground states are close to the pure quantum state through quantum state tomography (QST). Our method would be the basis for an almost ideal system corresponding to QA or QAOA using trapped ions. This method can also be applied to 2D ion crystals, which can rapidly scale up the number of ion qubits beyond the classical computational capabilities \cite{wang2020coherently,qiao2022observing}. 

\section{Results}

\subsection{Experimental setup scheme}

\begin{figure}
    \centering
    \includegraphics{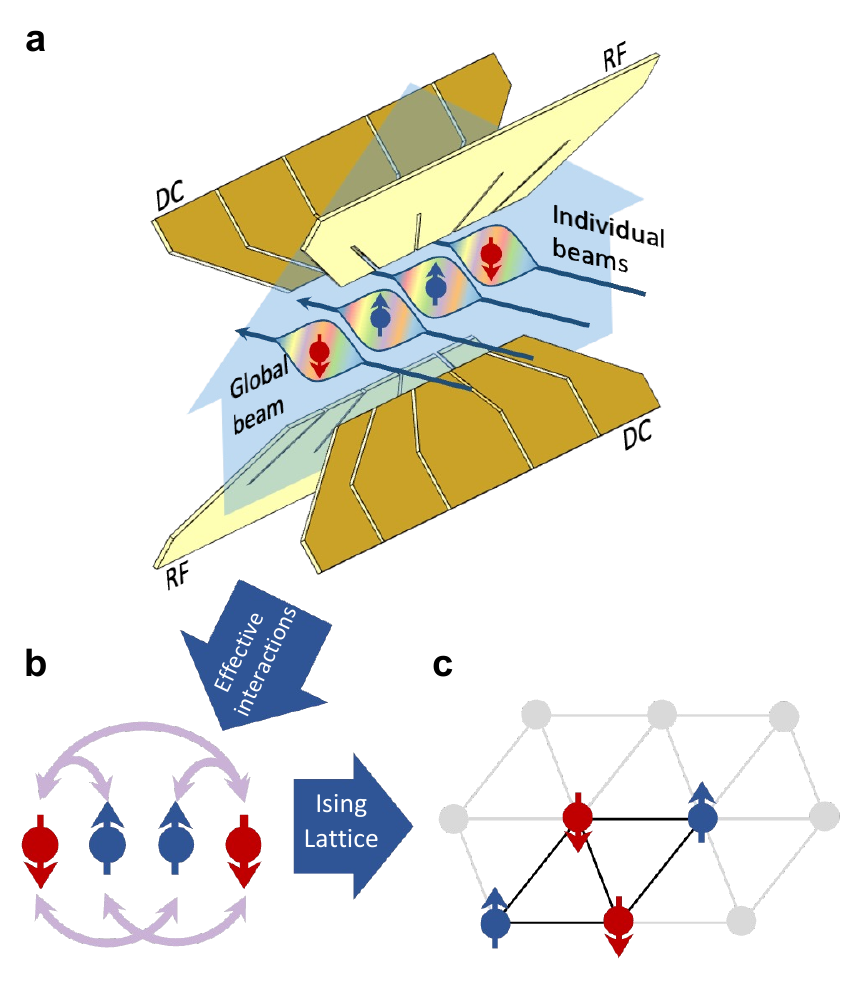}
    \caption{\label{fig:fig1} Experimental realization of programmable Ising models. \textbf{a}. Experimental setup. A chain of ions trapped in the blade trap is coherently manipulated via the stimulated Raman transition induced by two orthogonal laser beams. One of the Raman beams is divided into beams array, providing individual addressing capability on each ion. \textbf{b}. Example of effective spin-spin interactions. Arbitrarily connected interactions can be achieved by properly designing the modulation pattern of each individual beam. \textbf{c}. Example of mapping Ising lattice. Depending on the connectivity graph of the spin-spin interactions, it can be mapped to certain parts of the Ising lattice.}
\end{figure}

We employ a trapped-ion-based quantum simulator, which consists of up to four $\yb$ ions in a linear-chain configuration, as illustrated in \figref{fig:fig1}{a}. A single effective spin-1/2 is encoded within the ground manifold of each ion, denoted by $\ket{\downarrow} = {{^2}S_{1/2}}\ket{F = 0, m_F = 0}$ and $\ket{\uparrow} = {^2}S_{1/2}\ket{F = 1, m_F = 0}$, with an energy gap approximately $\omega_0 \approx 2\pi \times12.64~\mathrm{GHz}$ \cite{olmschenk2007manipulation}. All spins can be initialized to the $\ket{\downarrow...\downarrow}$ state by standard optical pumping, and their quantum states are measured via spin state-dependent resonant fluorescence. By utilizing the multi-channel photomultiplier tube (PMT) we can perform site-resolved detection on multiple ions \cite{chen2023scalable}. For more details on our quantum simulator, see the Method section. 

Coherent manipulations of spin states are achieved through stimulated Raman transitions. To enable the capability of individual control of each spin, we split one of Raman beams into a laser array and then let them pass through a multi-channel acoustic-optical modulator (AOM) \cite{lu2019global}. Effective spin-spin interactions can be engineered by coupling ions to their collective motional modes, and here we apply bi-chromatic fields with frequencies of $\omega_0 \pm \mu$, leading to a spin-dependent force (SDF) on the entire ion chain \cite{haljan2005spin},
\begin{equation} \label{equ:SDF}
H_\mathrm{SDF}(t) = \sum_{j,m}
    \dfrac{\eta_{j,m}\Omega_j}{2}(
        a_m \ce^{-\ci(\mu - \nu_m) t}\ce^{\ci \phi_j}
        + \mathrm{H.c.}
        )\pauli{x}^{(j)}.
\end{equation}
Here, $\pauli{x}^{(j)}$ represents the Pauli matrix of the $j$-th ion, $\eta_{j,m}$ denotes the scaled Lamb-Dicke parameter associated with $j$-th ion and $m$-th motional modes, $\nu_m$ corresponds to the frequency of the $m$-th motional modes, and $a_m$ is the corresponding annihilation operator. 

The earlier demonstrations of the Ising models within trapped-ion systems primarily concentrated on adiabatic regimes \cite{kim2010quantum,islam2011onset}. However, this approach exhibits certain limitations, notably the residual spin-motion couplings inevitably lead to decoherence of the system during long-time evolution. Also, the interaction graph has limited flexibility in programming. To address the above issues, we leverage the optimal control method to optimize the control parameters. This strategy not only allows us to eliminate undesired spin-motion couplings but also provides the means to program spin-spin interactions. Specifically, in our work, we primarily modulate the phases $\phi_j$ and amplitudes $\Omega_j$ of the SDF over time \cite{lu2019global}. Through carefully designed modulation patterns, we achieve an effective evolution operator as below,
\begin{equation} \label{equ:Ut}
U_\mathrm{Ising}\left( \left\{ \Theta_{j,j'} \right\} \right) = \mathrm{exp} \left[ - \ci \sum_{j<j'} \Theta_{j,j'} \pauli{x}^{(j)}\pauli{x}^{j'} \right].
\end{equation}
This operator is valid at a pre-determined time $\tau_g$. Here, the values of any $\Theta_{j,j'}$ can be arbitrarily tuned through the design of the modulation pattern. Thus, it offers the possibility of mapping \equref{equ:Ut} to an Ising lattice with arbitrary weighted connectivity, as depicted in \figref{fig:fig1}{b-c}.

\begin{figure*}[htbp]
    \centering
    \includegraphics[scale = 1]{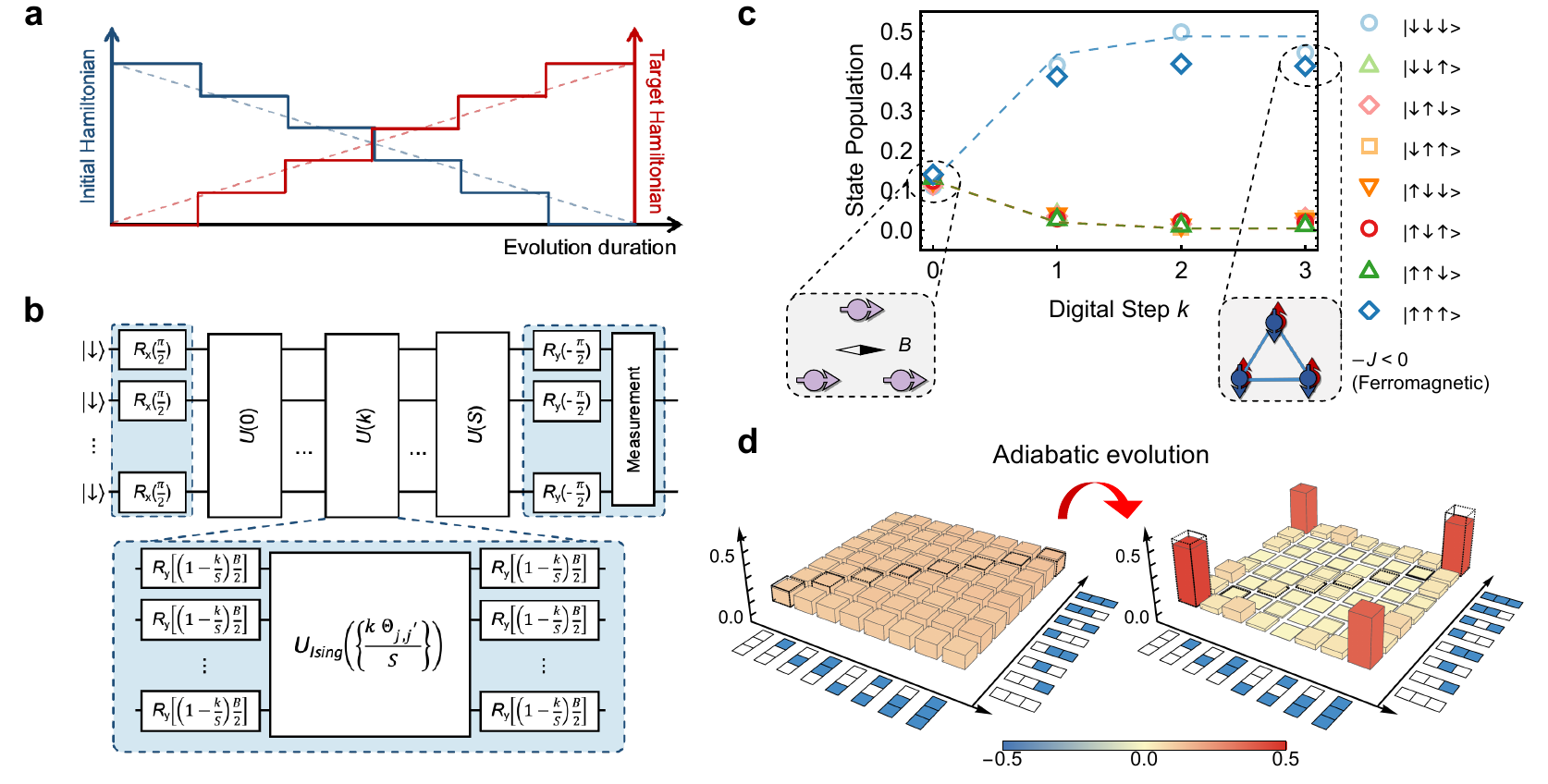}
    \caption{\label{fig:fig2} Digitized adiabatic evolution. \textbf{a}. Adiabatic evolution. In the conventional adiabatic evolution, the strengths of the initial (targeted) Hamiltonians are continuously ramped down (up) (shown in the dashed lines). For the digitized approach, the continuous ramping is discretized into several stages (shown in the solid lines). \textbf{b}. Circuits description of the digitized adiabatic evolution. After initializing the whole ion chain to $\ket{\downarrow}$ state, $\pi/2$- rotations are applied to prepare the ground state of the transverse field. The whole time evolution is decomposed into a series of single-spin rotations and multi-spin interactions. Note that the second-order Trotterization is employed in our experiments. \textbf{c}. Results of the states' population for each digital step. Here we choose the fully connected 3-spin ferromagnetic Ising model as an example. The open markers indicate experimental results, while the dashed lines guide the theoretical digital evolution. The error bars of one standard deviation are smaller than the size of the markers. \textbf{d}. Results of the state tomography. We show the real parts of the density matrices obtained in the experiments. The colored squares around axes indicate spin states, where the filled blue and empty squares represent $\ket{\uparrow}$ and $\ket{\downarrow}$ states, respectively. The left one indicates the initial ground state of the transverse field along the $\pauli{y}$-direction, with an initialization fidelity estimated to be 0.964(2). The right part corresponds to the final ground state of the ferromagnetic Ising model. The dashed boxes represent the diagonal components of the density matrices obtained from the ideal digitized evolution. All the error bars here and below represent one standard deviation.}
\end{figure*}

To evaluate our approach in simulating arbitrary Ising models, we employ the adiabatic evolution techniques to prepare their ground states. By first preparing all spins to the eigenstate of a transverse field, i.e. $H_\mathrm{ini} = -B\sum_{j}\pauli{y}^{j}$ ($B>0$), we could finally reach the ground state after slowly ramping up (down) the Hamiltonian of the Ising model (transverse field). As the programmed Ising model described in \equref{equ:Ut} is only valid at a discrete time of $\tau_g$, the digitized adiabatic evolution is employed here \cite{barends2016digitized}, as illustrated in \figref{fig:fig2}a. Within this framework, the time evolution is discretized into alternatively applied spin-spin interactions and transverse field, as shown in \figref{fig:fig2}b, and the strengths of the transverse field and the Ising interaction remain constant within each digitized step. This digitized evolution would approach the continuous one as the number of the digitized step $S$ is large enough. To further verify the final state obtained via the digitized adiabatic evolution, we execute QST circuits after the adiabatic evolution. Subsequently, we reconstruct the density matrices of the final states based on the results obtained from measurements in different bases.

\subsection{Ising models with three sites}

We begin our demonstration by focusing on the Ising models consisting of three spins. In this situation, three $\yb$ are confined in the trap, and the radial motional frequencies along $x$-direction are estimated as $\{ \nu_1, \nu_2, \nu_3 \} = 2\pi\times\{2.173, 2.118, 2.039\}~\mathrm{MHz}$. To simplify the experimental demonstration, we consider the scenarios where the interaction strength between any two spins is restricted to values of $-J, J$ or $0$ ($J>0$), corresponding to ferromagnetic, anti-ferromagnetic or no interactions, respectively. The three-site models can be organized into a one-dimensional (1D) line or a two-dimensional (2D) triangular configuration. Given that 1D cases which require only the nearest neighbor couplings have been extensively studied and yield ground states similar to GHZ states \cite{kim2010quantum}, we focus only on the exploration of the high-dimensional Ising models.

\begin{figure*}[htbp]
    \centering
    \includegraphics[scale = 1.]{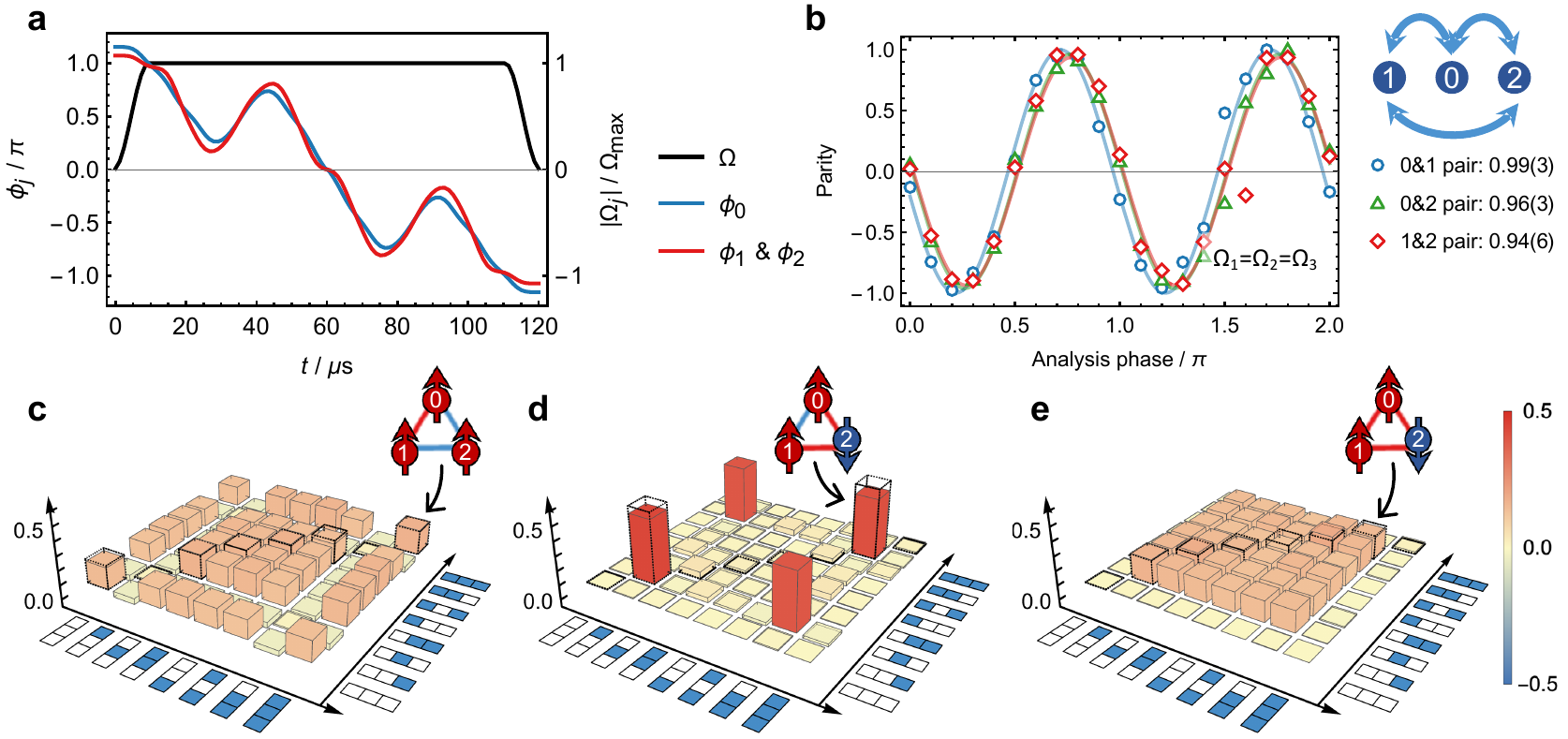}
    \caption{\label{fig:fig3} Experimental results of the prepared ground states for different 3-spin Ising models. \textbf{a}. Modulation pattern to realize 3-spin Ising model. In the pulse optimization, we set the frequency of the SDF to be $\mu = 2\pi \times 2.078$~MHz, and the operation duration $\tau_g$ to be 120~$\mu$s. The values of $|\Omega_j|$ (shown in the black curve) for all ions are set to be the same in the experimental optimization, while the sign of $\Omega_j$ can be utilized to adjust the sign of the spin-spin interactions. The phase modulation is performed smoothly and continuously (shown in blue and red curves), minimizing the noise in the AOM that occurs when the phase is suddenly changed. \textbf{b}. Characterization of spin-spin interactions. Utilizing the modulation pattern shown in \textbf{a}, we address each ion pair to prepare the corresponding maximally entangled state. We then measure the parity oscillations of these entangled states, with the sign of the oscillations indicating the sign of the interaction. The legend on the left side denotes the fitted contrasts. \textbf{c}-\textbf{e}. Real parts of the reconstructed density matrices for different Ising model ground states. In each subfigure's inset, a possible alignment of spins for the ground state is illustrated. Blue (red) lines connecting spins represent (anti-)ferromagnetic interactions.}
\end{figure*}

In the context of the 2D triangular lattice, when all interactions between spin pairs are ferromagnetic, the spin pairs between different sites naturally align in parallel, thereby minimizing the overall energy of the entire system. Consequently, the ground state of the system adopts the 3-GHZ state, as depicted in \figref{fig:fig2}{c-d}. Here, the digitized evolution is divided into three steps ($S=3$) to balance the Trotter error and the experimental operational errors. The modulation patterns for the phases $\phi_j$ and absolute amplitudes $|\Omega_j|$ of SDF are outlined in \figref{fig:fig3}a. Note that the amplitude modulation primarily serves as pulse shaping to eliminate errors arising from off-resonant coupling to the carrier transition. By setting the sign of all $\Omega_j$ to be the same, a ferromagnetic Ising model can be achieved. All the pairs of spin-spin interactions are experimentally characterized, as summarized in \figref{fig:fig3}{b}. In detail, we address targeted spin pairs, such as the $j$- and $j'$-th ions, to prepare the maximal entangling states, by utilizing designed modulation patterns with the proper amplitudes. After applying the analysis rotation, $\mathrm{exp}[\ci \pi (\sigma_{\phi}^{j}+\sigma_{\phi}^{j'})/4]$, to the prepared entangling state, the measured parity oscillation $P(\phi)$ would reveal the sign of $\Theta_{j,j'}$, as the relation of $P(\phi) \propto \mathrm{sign}(\Theta_{j,j'})\sin(2\phi)$.

In the digitized evolution, the values assigned to the transverse field $B$ and the spin-spin interaction $J$ are $\pi/2$ and $\pi/3$, respectively. The QST result yields a fidelity $\mathcal{F}$ of 0.835(4) for the prepared quantum ground state. Here $\mathcal{F}$ is defined as $(\mathrm{Tr}(\sqrt{\sqrt{\rho_\mathrm{th}}\rho_\mathrm{exp}\sqrt{\rho_\mathrm{th}}}))^2$, where $\rho_\mathrm{th}$ and $\rho_\mathrm{exp}$ are the density matrices obtained from the simulation of ideal digitized evolution and the experimental QST data, respectively. The observed errors primarily stem from the depolarization during effective spin-spin interactions. This point can be verified that if we only consider the diagonal parts of the density matrices ($\rho^\mathrm{dig} = \mathrm{Diag}\left[\rho\right]$), the overlap between the experimental result and the theoretical value, which is denoted as $\mathcal{F}_\mathrm{diag} = (\mathrm{Tr}(\sqrt{\sqrt{\rho^\mathrm{diag}_\mathrm{th}}\rho^\mathrm{diag}_\mathrm{exp}\sqrt{\rho^\mathrm{diag}_\mathrm{th}}}))^2$, increases to 0.944(7). This also suggests that improving the fidelity of spin-spin interactions can further enhance the fidelity of the prepared ground state. 

Upon exploring the pure ferromagnetic Ising lattice, we proceed to modify partial interactions to be antiferromagnetic, subsequently preparing the corresponding ground states. The results are summarized in \figref{fig:fig3}{c-e}. A crucial observation arises when one or all three bonds are changed to anti-ferromagnetic interactions. In these scenarios, the spin pairs are unable to hold the lowest energy simultaneously due to the conflict of the interactions, resulting in geometric frustration \cite{kim2010quantum}. This phenomenon leads to an augmentation in ground state degeneracy, with six degenerate ground states present for both the single bond and all bonds antiferromagnetic Ising lattices, as illustrated in \figref{fig:fig3}{c-d}. The fidelities ($\mathcal{F}$) of the prepared ground states are estimated as 0.858(5) and 0.871(5) for the single bond and all bonds anti-ferromagnetic Ising lattice, respectively, while the overlap of the diagonal parts ($\mathcal{F}_\mathrm{diag}$) are 0.984(7) and 0.972(7). Except for the aforementioned cases, when two bonds are flipped to anti-ferromagnetic, the corresponding ground state reverts to a GHZ-like state. This is because all spin pairs can simultaneously assume their lowest energy states, as displayed in \figref{fig:fig3}{e}. The state fidelity ($\mathcal{F}$) and diagonal overlap ($\mathcal{F}_\mathrm{diag}$) for this scenario measure at 0.844(5) and 0.934(7), respectively.

Note that, the 3-spin Ising model demonstrated above all utilize the same phase modulation patterns. For the Ising model in \figref{fig:fig3}{d}, we reverse the sign of amplitude $\Omega_2$ to make $\Omega_0 = \Omega_1 = -\Omega_2$, allowing us to construct the Ising model of $-\pauli{x}^{0}\pauli{x}^{1} + \pauli{x}^{0}\pauli{x}^{2} + \pauli{x}^{1}\pauli{x}^{2}$. In the Ising model presented in \figref{fig:fig3}{c}, we also set $\Omega_0 = \Omega_1 = -\Omega_2$, and in fact the experimentally prepared state is the highest excited state. Nevertheless, this state is equivalent to the ground state of the Ising model $\pauli{x}^{0}\pauli{x}^{1} - \pauli{x}^{0}\pauli{x}^{2} - \pauli{x}^{1}\pauli{x}^{2}$ except for certain phase sign differences, which can be corrected in the density matrix reconstruction. Similarly, for results showcased in \figref{fig:fig3}{e}, we also prepare the highest excited state of the ferromagnetic Ising model.

\subsection{Ising models with four sites}

\begin{figure*}[htbp]
    \centering
    \includegraphics[scale = 0.99]{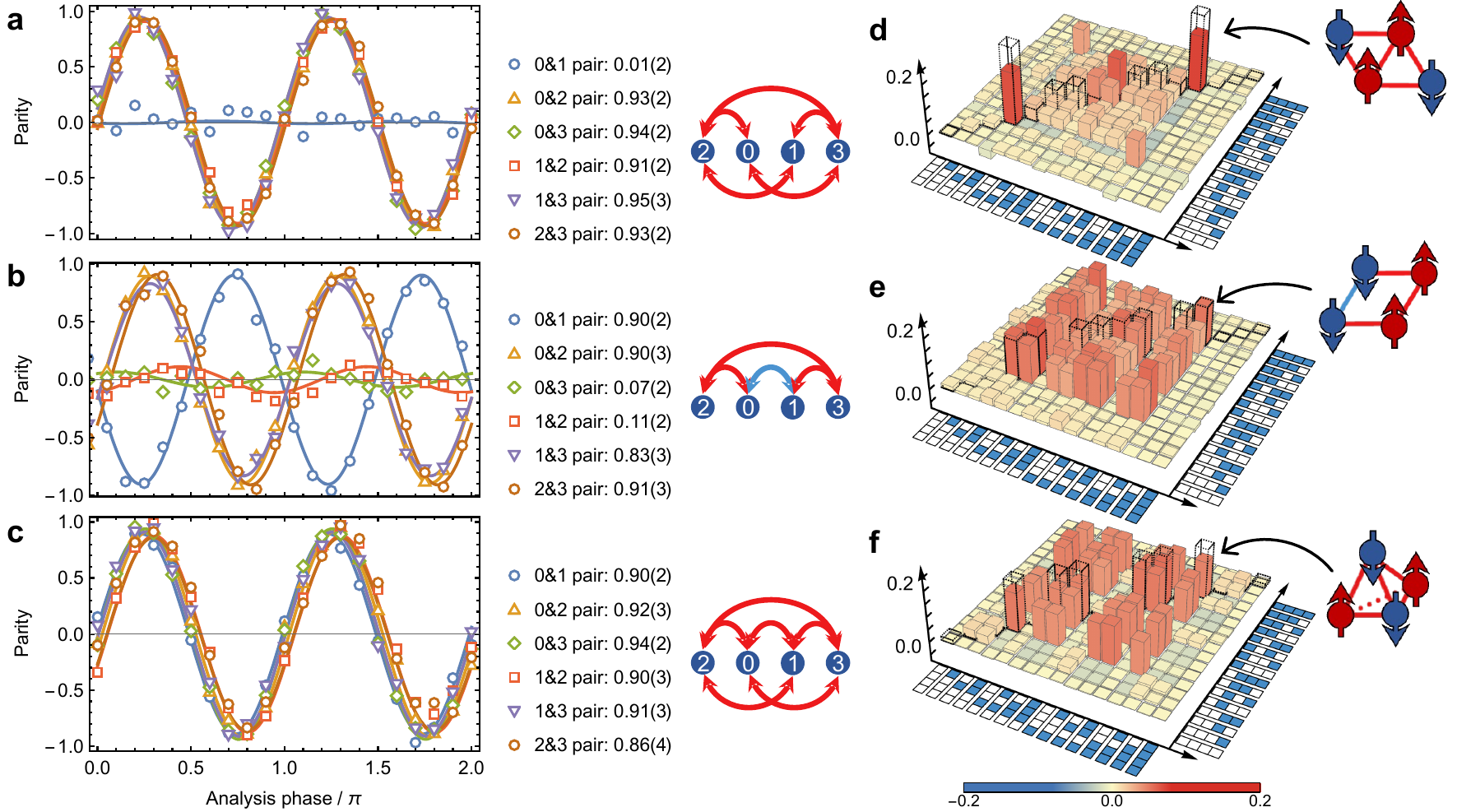}
    \caption{\label{fig:fig4} Experimental results of prepared ground states for 4-spin Ising models.
    All the spin pairs in each Ising model are addressed to characterize the sign of the spin-spin interactions, as depicted in \textbf{a} through \textbf{c}.
    In \textbf{a}, we can clearly observe a near cancellation of interaction between the spin pair 0-1. While in \textbf{b}, the interactions for both spin pairs 0-3 and 1-2 are simultaneously canceled, and the interaction between the pair 0-1 is reversed to be ferromagnetic as well.
    Then \textbf{c} presents a typical representation of fully connected anti-ferromagnetic interactions, which can be treated as a three-dimensional Ising lattice. In \textbf{d} through \textbf{f}, we provide the experimentally obtained density matrices resulting from digitized adiabatic evolution, for the interaction configurations depicted in \textbf{a} through \textbf{c}, respectively}
\end{figure*}

We now transition our focus to the Ising models consisting of four spins. Towards these models, we introduce an additional ion to the trap, increasing the total number of ions to four. The trap frequencies are measured as $\{ \nu_1, \nu_2, \nu_3, \nu_4 \} = 2\pi\times\{2.179, 2.138, 2.081, 2.005\}~\mathrm{MHz}$. In order to mitigate Trotterization errors, the digitized step is heightened to four as well. The inclusion of an extra spin facilitates an extension of 2D triangular structure shown in \figref{fig:fig3}{} to a larger lattice. To illustrate, the anti-ferromagnetic triangular lattice showcased in \figref{fig:fig3}{e} is extended to a bi-triangular lattice, depicted in \figref{fig:fig4}{a} and \figref{fig:fig4}{d}. Evidently, the contrast in parity oscillation between spins 0 and 1 almost disappears, signifying the absence of spin-spin interaction in this particular pair. This newly introduced site reduces the count of degenerate ground states to two, compelling both triangular sub-structures to maintain a symmetrical alignment of spins. Experimentally, we achieve the ground state with a fidelity $\mathcal{F}$ of 0.576(3), accompanied by a diagonal overlap $\mathcal{F}_\mathrm{diag}$ of 0.853(5). In comparison to the results from three-site situations, the fidelity of the prepared state diminishes due to the increase of the digital steps and escalated depolarization error during the spin-spin interactions.

We also attempt to realize a square lattice by simultaneously removing the interactions between spin pairs of 0-3 and 1-2, as shown in \figref{fig:fig4}{b} and \figref{fig:fig4}{e}. Furthermore, the interaction between spin pair 0-1 is engineered to be ferromagnetic, while the remaining interactions retain anti-ferromagnetic. This configuration also manifests high geometric frustration within its ground state. The experimental result reveals a fidelity $\mathcal{F}$ of 0.750(4), while the diagonal overlap $\mathcal{F}_\mathrm{diag}$ approximates to be 0.930(4).

An extension into a three-dimensional (3D) Ising lattice with a tetrahedral structure can be realized through the all-to-all connection, exemplified in \figref{fig:fig4}{c-f}. This structure constitutes a fundamental and pivotal component in various solid crystals, typified by pyrochlore oxides. The magnetic characteristics of this configuration have garnered considerable theoretical and experimental attention over the last few decades. In this scenario, we prepare the ground state of a solitary tetrahedral lattice featuring antiferromagnetic interactions, thereby inducing geometric frustration once more. All the ground states exhibit a consistent feature: the presence of two spins up and two spins down, mirroring the ice rule (two-in, two-out) in the spin-ice system. The fidelity of the ground state we prepare measures at 0.719(3), while the diagonal overlap rests at 0.886(4). The modulation patterns for all the 4-spin Ising models can be found in the Methods section.

All the above results are corrected to remove the detection error (Methods). Meanwhile, we can briefly estimate error contributions in our demonstration. Our experiments are conducted using the same setup detailed in a previous study \cite{lu2019global}, and as such, the sources of error remain consistent with those outlined in that work. As demonstrated in that study, the errors associated with 3-spin and 4-spin operations can be estimated to be approximately 5\% and 7\%, respectively. Consequently, for the 3-spin Ising models, our three-step digitized evolution would yield total fidelities at approximately the 85\% level. Similarly, for the 4-spin models, a four-step evolution would yield total fidelities around the 75\% level. Given these error estimates, it becomes evident that further improvement in multi-spin operations is crucial to enhance the performance of programmable Ising models.

\section{Conclusion and outlook}
In conclusion, our study brings an efficient approach to realize programmable Ising models featuring arbitrary connectivity, by leveraging the all-to-all capability among the trapped ions and the individual controllability. We suggest that our approach offers not only efficient implementation but also remarkable scalability. Recent research has demonstrated that the optimization of modulation patterns required for achieving programmable Ising interactions demands only polynomial computational resources \cite{grzesiak2020efficient}, thus affirming the scalability of our approach. Through the integration of programmable Ising interactions with digitized adiabatic evolution, we have experimentally achieved the successful preparation of quantum ground states for various Ising lattices comprising up to four spin sites. These results mark a significant step in the realization of practical QA and QAOA algorithms on NISQ devices based on trapped-ion platforms.

For future research, the studies on more complex spin systems that own i.e. XY or Heisenberg interactions should included. Generalized spin models with arbitrary connectivity open up significant opportunities for exploring richer quantum phenomena. Moreover, the digitized adiabatic evolution employed here can be accelerated by implementing the techniques of shortcuts to adiabaticity (STA) \cite{hegade2021shortcuts}. A recent study also suggests a mechanism of self-healing in digitized adiabatic state preparation \cite{kovalsky2023self}. It suggests that the accumulation of infidelity during digitized adiabatic evolution could scale lower than the bounds given by general Trotter errors. Both techniques and features hold the promise of further enhancing the computational power of programmable trapped-ion quantum simulators, paving the way for novel advancements in near-term quantum computation and simulation.  During the preparation of this paper, we became aware of related studies about engineering programmable spin models by utilizing multi-tone drives \cite{wu2023qubits, shapira2023programmable}.

\bibliography{IsingLattice}

\setcounter{figure}{0}
\renewcommand{\figurename}{Extended Data Fig.}

\section{Methods}

\subsection{Experimental setup and detection error correction}

\begin{figure*}[htbp]
    \centering
    \includegraphics[scale = 0.99]{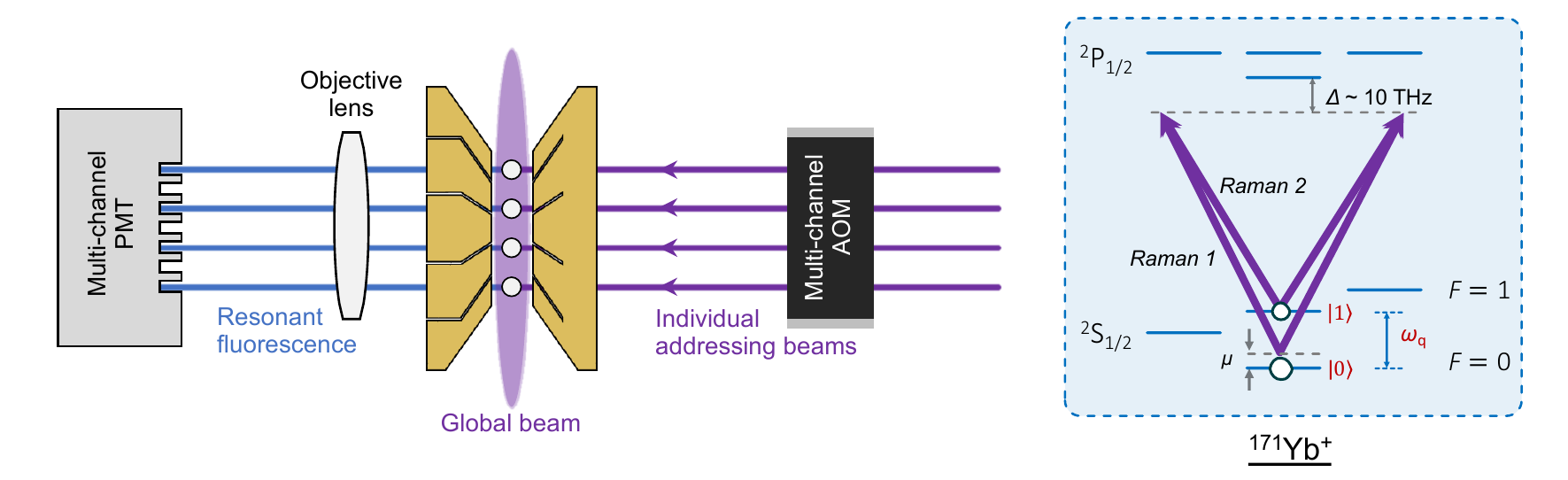}
    \caption{\label{fig:exfig1} Experimental setup. The main components for our programmable quantum simulator are illustrated. The inset figure shows the energy levels of the $\yb$ ion.}
\end{figure*}

In \exfigref{fig:exfig1}, we provide a detailed view of our experimental setup, highlighting both the individual laser addressing system for independent ion control and the imaging system for spin-state readout. Our experimental apparatus employs a pulsed laser (COHERENT Mira HP, not shown in the figure) for stimulated Raman transitions with a center wavelength of 377 nm. This laser is detuned by approximately 10 THz from the transition frequency between the $^2S_{1/2}$ and $^2P_{1/2}$ levels. The pulsed laser has a pulse width of about 4 ps and operates at a repetition rate of around 76 MHz.

To drive the stimulated Raman transition, the laser beam is divided into two separate beams. These beams propagate in equal optical paths and overlap at the ion-chain position with orthogonal wave vectors. Specifically, one of the laser beams is divided into a beam array using a diffractive optical element (from Holo.Or). This beam array passes through a multi-channel Acousto-Optic Modulator (AOM) (Gooch \& Housego AOMC 220-5) and is then tightly focused onto the ion chain. Both ion spacing and beam spacing are adjusted to be approximately 5 µm. It's worth noting that our system has the capability to independently control a maximum of five ions, although, for our experimental demonstration, we employ up to four ions.

\begin{figure*}[htbp]
    \centering
    \includegraphics[scale = 0.99]{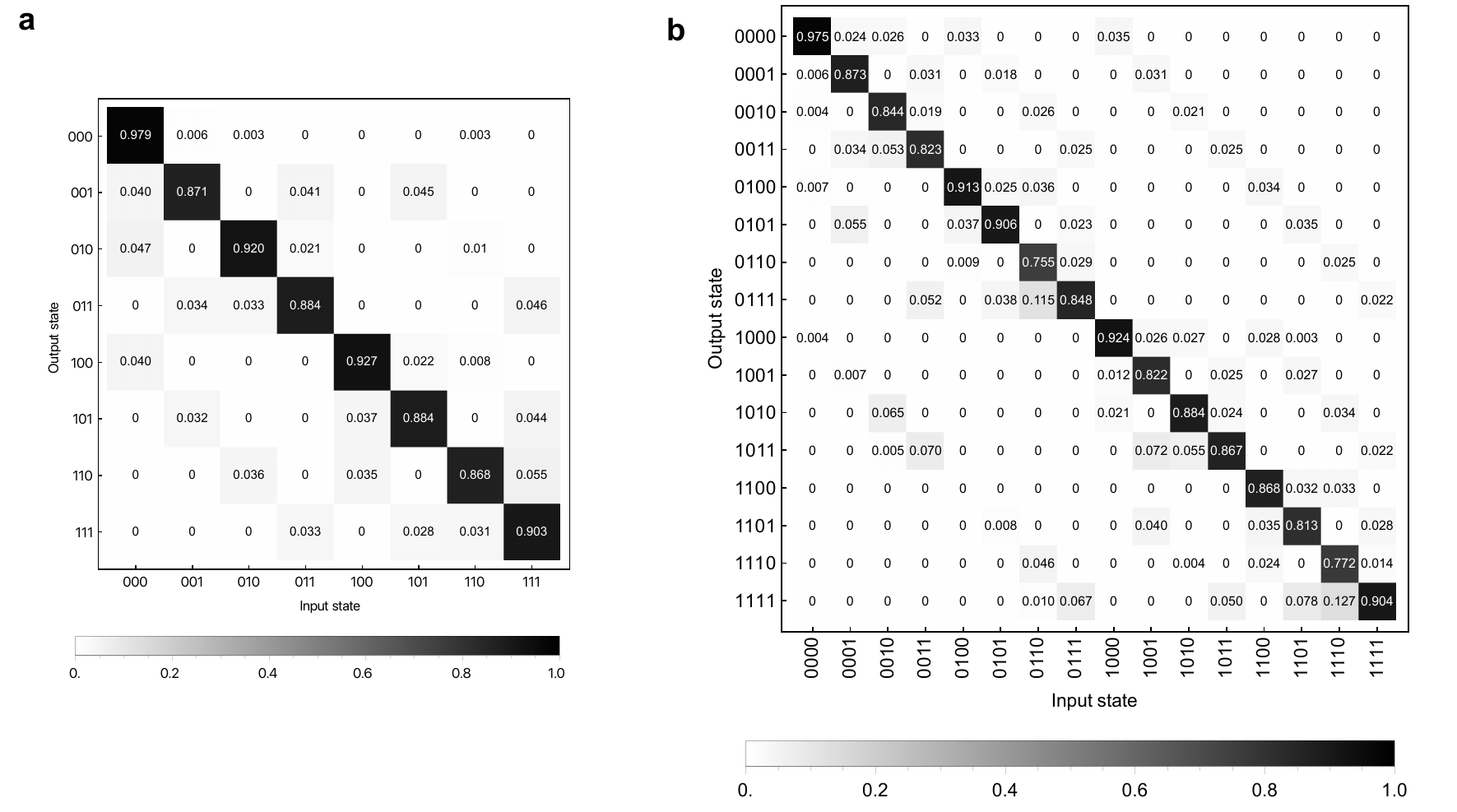}
    \caption{\label{fig:exfig2} Detection error matrices. The average detection fidelity for the three-ion system is estimated to be 90.4\%, while for the four-ion system, it is estimated to be 86.2\%. These value lead to single-ion dectection fidelities of around $96.7\%$ ($={\sqrt[3]{90.4\%}}$) for the three-ion system and $96.3\%$ ($={\sqrt[4]{86.2\%}}$) for the four-ion one. } 
\end{figure*}

During the detection process, fluorescence photons that are emitted by the whole ion chain are collected by an objective lens with a numerical aperture (N.A.) of 0.6 (Photon Gear 15470-S) and then counted using a linear array PMT module (Hamamatsu H11460). To achieve spatial-resolved measurements, the nearest ions are mapped to the next-nearest channels of the PMT. The total detection duration lasts approximately 250 µs. However, our detection process faces certain challenges. The low quantum efficiency of the PMT, approximately 20\% at 370 nm, along with significant crosstalk between nearby ions, limits our detection fidelity for multiple ions. To characterize the detection error for three-ion and four-ion cases, we prepare serials of input quantum states and then measure the output state distribution. The experimentally obtained error matrices are provided in \exfigref{fig:exfig2}.

Both error matrices reveal a single-ion detection fidelity of approximately 96.5\%. The infidelity includes the errors stemming from state leakage due to off-resonant coupling during detection (around 2\%) and also the crosstalks between nearby ions (around 1.5\%). With these error matrices, we can manually correct the raw detection results, denoted by $\mathbf{P}^\mathrm{meas} = \{ p^\mathrm{meas}_{0...0},...,p^\mathrm{meas}_{1...1} \}$, to remove these errors. In practice, we can estimate the real state population using the reverse of the error matrix $\mathbf{M}$, which allows us to obtain $\mathbf{P}^\mathrm{real} = \mathbf{M}^{-1}.\mathbf{P}^\mathrm{meas}$, where $\mathbf{P}^\mathrm{real}$ is the estimated real state population. To avoid non-physical outcomes from direct inversion, we apply the maximum-likelihood method. This method estimates the real state population by minimizing the 2-norm of $\| \mathbf{P}^\mathrm{meas} - \mathbf{M}.\mathbf{P}^\mathrm{real} \| _2$.

\subsection{Optimization method for pulse modulation}
\begin{figure*}[htbp]
    \centering
    \includegraphics[scale = 0.99]{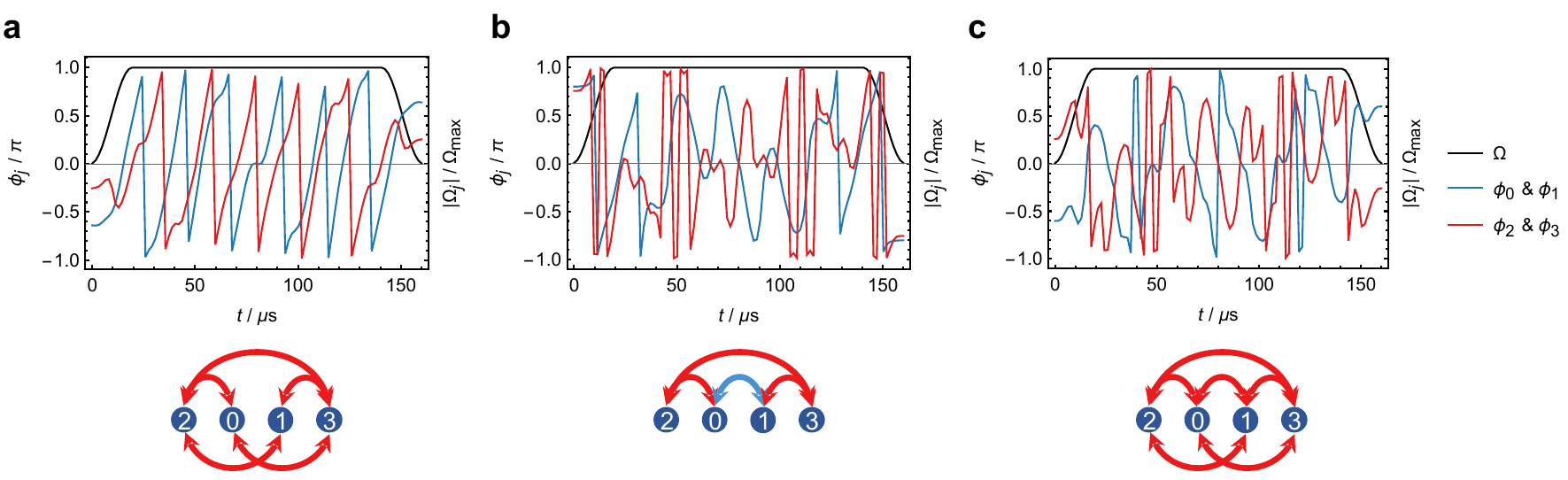}
    \caption{\label{fig:exfig3} 
    Modulation patterns for four-spin Ising models.
    In the experiments, we set the value of $\mu$ to be $2\pi\times 2.043$ MHz for all the patterns. Given the symmetry of the Ising models, the phase modulation patterns are identical for the center two ions and the outer two ions. It's worth noting that any seeming discontinuities in the phase modulation pattern are the results of the modulo $2\pi$ effect. The number of segments in all four-ion cases is 100.} 
\end{figure*}

The optimization method employed here to obtain the modulation patterns is similar to the approach utilized in the prior work\cite{lu2019global}. When integrating the SDF Hamiltonian in \equref{equ:SDF} over time, we obtain an evolution operator given by:
\begin{eqnarray}
    U(t) = \mathrm{exp} 
        \left[
            \sum_{j,m} (\alpha_{j,m}(t) a_m^\dagger - \alpha_{j,m}^\star(t) a_m) \pauli{x}^{(j)}  \right. \\ \nonumber
        \left.  
            - \ci \sum_{j<j'} \theta_{j,j'}(t) \pauli{x}^{j}\pauli{x}^{j'}
        \right].
\end{eqnarray}
Here, $\alpha_{j,m}$ represents the trajectories of the $m$-th motional modes in the phase space contributed by the $j$-th ion, while $\beta_{j,j'}$ indicates the effective two-body interaction between the $j$-th and $j'$-th ions. The analytical forms of $\alpha_{j,m}$ and $\theta_{j,j'}$ are given as follows:
\begin{equation}
    \alpha_{j,m}(\tau) = - \ci \eta_{j,m} \int_{0}^{\tau} \dfrac{\Omega_j(t)e^{-\ci \phi_{j}(t)}}{2} e^{\ci(\nu_m - \mu)t} dt,
\end{equation}
\begin{eqnarray}
    \theta_{j,j'}(\tau) &=& - \sum_{m} \int_{0}^{\tau} dt_2 \int_{0}^{t_2} dt_1 \dfrac{\eta_{j,m} \eta_{j',m} \Omega_j(t_2)\Omega_{j'}(t_1)}{2} \nonumber \\ 
    &&\sin \left[
            (\nu_m - \mu)(t_2 - t_1) - (\phi_j(t_2) - \phi_{j'}(t_1))
        \right].  \nonumber \\
\end{eqnarray}
The target for optimization is minimizing the cost function $\sum_{j,m}|\alpha_{j,m}(\tau_g)|$ for a given $\tau_g$, while the set of $\{ \theta_{j,j'}(\tau_g) = \Theta_{j,j'} \}$ yields constraints for optimization. In practice, we discretize the phase $\phi_j$ and amplitude $\Omega_j$ into $K$ segments, each with a segment duration of $\tau_s = \tau/K$. In contrast to the modulation patterns shown in the previous work \cite{lu2019global}, in this work, we opt for a larger number of segments exceeding a hundred. For instance, in the case of the three-ion system presented in \figref{fig:fig3}{a}, we choose a value of $K$ equal to 100. We now provide the modulation patterns for the four-spin Ising models achieved in our experiments, as shown in \exfigref{fig:exfig3}.

\

\textbf{Data availability}: 
  The data that support the findings of this study are available from the authors upon request.
\

\textbf{Acknowledgements}:
This work was supported by the Innovation Program for Quantum Science and Technology under Grants No. 2021ZD0301602, and the National Natural Science Foundation of China under Grants No.92065205, No.11974200, and and No.62335013. Y. L. also acknowledges support from the National Science Foundation of China under Grants No.~12004165, the Guangdong Basic and Applied Basic Research Foundation under Grant No. 2022B1515120021, and the Shenzhen Science and Technology Program under Grants No.~RCYX20221008092901006.
\

\textbf{Author contributions}:
    Y.L., W.C., S.Z. and K.Z. developed the experimental system conceived the project. Y.L. and J.-N.Z. optimized experimental schemes. Y.L. took and analysed the experimental data. K.K. supervised the project. All authors contributed to write the manuscript.
\
  
\textbf{Competing interests}:
    The authors declare that there are no competing interests.
\

\textbf{Author information}:
    Correspondence and requests for materials should be addressed to Y.L. and K.K.

\end{document}